\documentclass[doublecol]{epl2}
\usepackage{amsmath}
\usepackage{graphicx}
\usepackage{color}
\usepackage[normalem]{ulem}	

\usepackage{amssymb,amsfonts,amsmath}

\usepackage{subfigure}
\usepackage{hyperref}
\usepackage{epstopdf}
\newif\ifpdf
\ifx\pdfoutput\undefined
   \pdffalse
\else
   \pdfoutput=1
   \pdftrue
\fi
\ifpdf
   \usepackage{graphicx}
   \usepackage{epstopdf}
\else
   \usepackage{graphicx}
\fi
\def\half{{\textstyle \frac{1}{2}}}

\def\Tr#1{{\rm Tr}\left( #1 \right)}
\def\Det#1{{\rm Det}\!\left( #1 \right)}

\title{Exactly isochoric deformations of soft  solids}
 
\author{John S Biggins\inst{1}, Z Wei\inst{2},L Mahadevan\inst{2,3}}
\institute{ \inst{1} Cavendish Laboratory, JJ Thomson Ave, University of Cambridge, Cambridge, CB3 0HE, United Kingdom \\
\inst{2} School of Engineering and Applied Sciences, Harvard University, Cambridge, MA 02138, USA\\
\inst{3}Department of Physics, Harvard University, Cambridge, MA 02138, USA\\
\email{lm@seas.harvard.edu}}
 

\abstract{Many  materials of contemporary interest,  such as  gels, biological tissues and elastomers, are easily deformed but essentially incompressible. Traditional linear theory of elasticity implements incompressibility only to first order and thus permits some volume changes, which become problematically large even at very small strains. Using a mixed coordinate transformation originally due to Gauss, we enforce the constraint of isochoric deformations exactly to develop a linear theory with perfect volume conservation that remains valid until strains become geometrically large. We demonstrate the utility of this approach by calculating the response of an infinite soft isochoric solid to a point force that leads to a nonlinear generalization of the Kelvin solution. Our approach naturally generalizes to a range of problems involving  deformations of soft solids and interfaces in 2 dimensional and axisymmetric geometries, which we exemplify by determining the solution to a distributed load that mimics muscular contraction within the bulk of a soft solid.}
 
\begin{document}

\maketitle

A basic question in the theory of elasticity, particularly within the small-strain limit of linear elasticity, is the response function of an infinite solid to a point force, the Green's function of the solid known as the Kelvin solution \cite{kelvin1848displacement}. Knowledge of this function allows one to calculate the response of the solid to an arbitrary force distribution as the weighted sum of point force responses. In the presence of constraints such as incompressibility that are commonly used to approximate the mechanical behavior of many soft solids such as gels, elastomers and biological tissues, we can ask how this solution changes.  In this letter we answer this question by constructing a theory that is exactly isochoric and use it to construct the isochoric analog of the Kelvin point force solution that only implements the constraint of incompressibility to first order.  Although we focus on a point force in an infinite body for ease of exposition, our approach is easily generalizable to any 2-D or axisymmetric situation. 

\begin{figure*}\centering
\subfigure[Representations of Deformations]{\label{targandref}\includegraphics[width=1.0 \textwidth]{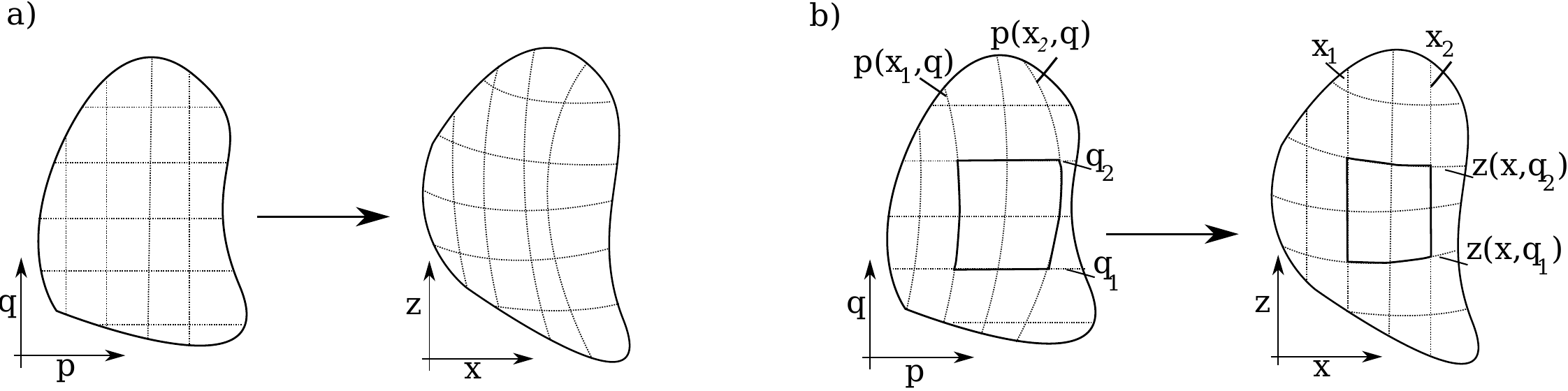}}
\caption{A two dimensional elastic body labeled by the coordinates $(p,q)$ is deformed into the target state with coordinates $(x,z)$. Top diagram shows lines of constant $p$ and $q$ in both states, the traditional representation of an elastic deformation. Lower diagram shows lines of constant $x$ and $q$ in both states, a representation of the deformation that makes volume preservation easy to implement. A patch of material defined by $x_1<x<x_2$ and $q_1<q<q_2$ is drawn in both states.}\label{fig:flatplates}
\end{figure*}

Soft solids often have bulk moduli several orders of magnitudes higher than their shear moduli. If a material is subject to a displacement field $\mathbf{u}$, the local shape change is described by the deformation-gradient $F=I+\nabla \mathbf{u}$, and the local volume changes by a factor of $\Det{F}$. The simplest  energy function for describing such a material is the compressible neo-Hookean model which, in 2-D, is
\begin{align}
E_{nl}=\frac{1}{2}\mu \left(\frac{\Tr{F\cdot F^T}}{\Det{F}}-2\right)+\frac{1}{2}B \left(\Det{F}-1\right)^2\label{nlenergy}
\end{align}
where $\mu$ and $B$ are the shear and bulk moduli. In equilibrium, balancing the gradients of the shear and bulk terms with respect to $\mathbf{u}$ implies that $\mu \nabla \mathbf{u} \sim B (\Det{F}-1)$. Hence for nearly incompressible materials (with $\mu/B\ll1$) we see that the volumetric strain is much lower than the shear strain, so that the bulk-modulus energy term becomes negligible. Then, the effective energy density has a contribution only due to shear deformations while the bulk modulus term enforces the constraint of incompressibility:
\begin{equation}
E=\half \mu (\Tr{F\cdot F^T}-2),\hspace{3em} \Det{F}=1\label{nlenergyconstrained}.
\end{equation}
The constraint $\Det{F}=1$ is then naturally conjugate to a pressure field $P$ that serves as a Lagrange multiplier. Then, in the presence of an external force distribution $\mathbf{f}$ the total  effective energy density is given by
\begin{equation}
\tilde{E}=\half \mu \Tr{F\cdot F^T}+P(\Det{F}-1)-\mathbf{f}\cdot \mathbf{u}\label{nlenergyincom}.
\end{equation}
where the last two terms correspond to the work done by these volumetric forces. Minimizing the total energy $\int \tilde{E} \mathrm{d}{\bf x}$ with respect to  $\mathbf{u}$ and $P$ yields the Euler-Lagrange equations
\begin{equation}
\nabla \cdot (\mu F-P F^{-T})=-\mathbf{f} \hspace{3 em} \Det{F}=1,\label{tradnleqns}
\end{equation}
where we identify $\mu F- P F^{-T}$ as the first Piola-Kirchhoff (PK1) stress-tensor.  The nonlinear constraint of volume preservation makes the equations rarely tractable analytically, unless various symmetries are imposed. However, in the small strain limit when  
 $\nabla \mathbf{u}$ and $\nabla P$ are small, we can linearize both  equations, so that in the limit $|\nabla {\bf u}| \ll 1$, we get
\begin{equation}
\mu \nabla ^{2} \mathbf{u}-\nabla P=-\mathbf{f}, \hspace{3 em} \nabla\cdot\mathbf{u}=0,\label{tradlinear}
\end{equation}
which are mathematically equivalent to Stokes equations for creeping viscous flow, modulo the interpretation of $\bf u$ as the velocity field, and $\mu$ as the dynamic viscosity. In a two-dimensional Cartesian coordinate system $p-q$, setting $\mathbf{u}=\hat{\mathbf{q}}\partial_p \alpha-\hat{\mathbf{p}}\partial_q \alpha$, where $\alpha(p,q)$ is the stream function automatically guarantees linearized  incompressibility $\nabla \cdot \mathbf{u}=0$. Then Eqn.\ (\ref{tradlinear}) reads as 
\begin{equation}
\mu \left(\begin{array}{cc}
-\alpha_{qqq}-\alpha_{ppq}\\
\alpha_{qqp}+\alpha_{ppp}
\end{array}\right)-\left(\begin{array}{cc}
P_p\\
P_q
\end{array}\right)=-\mathbf{f}\label{tradelasteqn},
\end{equation}
where subscripts denotes differentiation, i.e. $(\cdot)_x = \partial (\cdot)/\partial x$ etc. In a region with $\mathbf{f}=0$ eliminating $P$ reveals that $\alpha$ is biharmonic so that $\alpha_{qqqq}+2\alpha_{ppqq}+\alpha_{qqqq}=0$. For a point force $\mathbf{f}=f \mathbf{\hat{q}}$ at the origin, these equations were solved by Kelvin \cite{kelvin1848displacement} and yield:
\begin{align}
\alpha&=-\frac{f p \log\left(q^2+p^2\right)}{8 \pi  \mu },\hspace{1em} P=P_0+\frac{f q}{2 \pi   \mu(q^2+p^2) } \label{2dsol}.
\end{align}
Such traditional incompressible elastic solutions satisfy linear incompressibility,  $\nabla\cdot \mathbf{u}=0$,  not $\Det{F}=1$. Thus the resultant fields cannot  be substituted  into eq.\ (\ref{nlenergyconstrained}), so  linear solutions cannot be used as a starting point for even weakly non-linear calculations. Indeed, if one substitutes the linear result back into the full almost incompressible energy, eqn.\ (\ref{nlenergy}), whose behavior we are ultimately interested in, we find that the dilatational energy term  is
\begin{equation}
\half B(\Det{F}-1)^2=\half B(\nabla \cdot \mathbf{u}+\Det{\nabla \mathbf{u}})^2.
\end{equation}
If $\mathbf{u}$ is the usual incompressible linear solution the quadratic  term vanishes, but the quartic term, $\half B \left(\Det{ \nabla \mathbf{u}}\right)^2\sim B (\nabla \mathbf{u})^4$, does not. This erroneous contribution to the energy will become problematic when it is commensurate in size with the leading order quadratic shear term $\half \mu \Tr{\nabla \mathbf{u}\cdot \nabla \mathbf{u}^T}\sim\mu(\nabla \mathbf{u})^2$, which will occur when the strain approaches $\nabla \mathbf{u}\sim \sqrt{\mu/B}$. For an almost incompressible material  ($B/\mu\gg1$) the  solution produces unacceptably large volume changes at very small strains, and, in the limit of true incompressibility,  it is only valid for asymptotically small strains. In contrast, if $B\sim\mu$, the linear regime works well until $\nabla \mathbf{u}\sim 1$, at which point non-linear geometric effects guarantee that any linear theory must break down. This  will be an issue if we wish to use the traditional linear solution as a seed or a  boundary condition for a full finite element calculation utilizing eqn.\ (\ref{nlenergy}), or if we wish to use it to use it as a test function for the constrained energy, eqn. (\ref{nlenergyconstrained}).

\begin{figure*}\centering
\subfigure[Exact Volume Preservation]{\label{streamline}\includegraphics[width=0.24\textwidth]{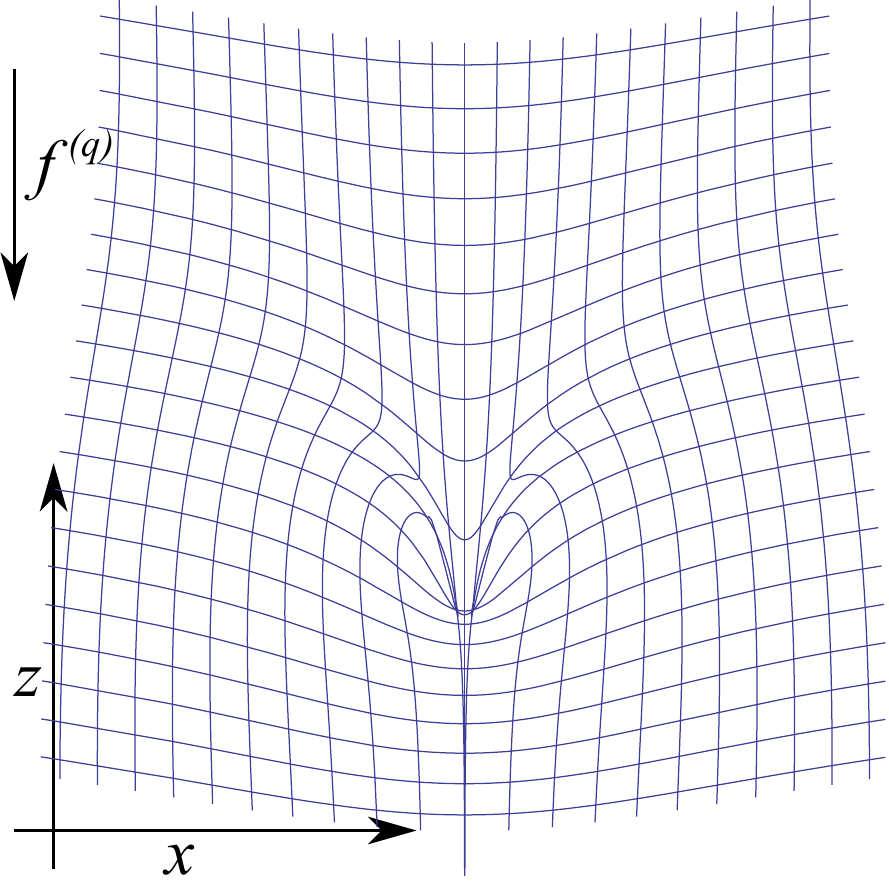}}
\subfigure[Conventional Kelvin Solution]{\label{linear}\includegraphics[width=0.24\textwidth]{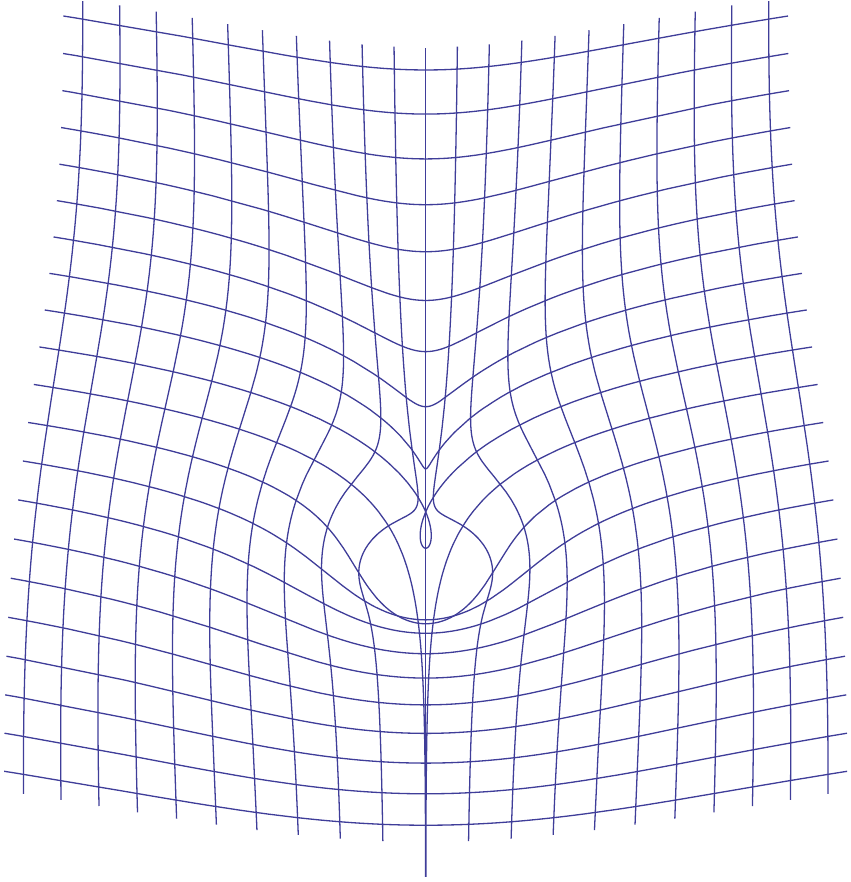}}
\subfigure[Volume changes in Kelvin solution]{\label{linearvol}\includegraphics[width=0.24\textwidth]{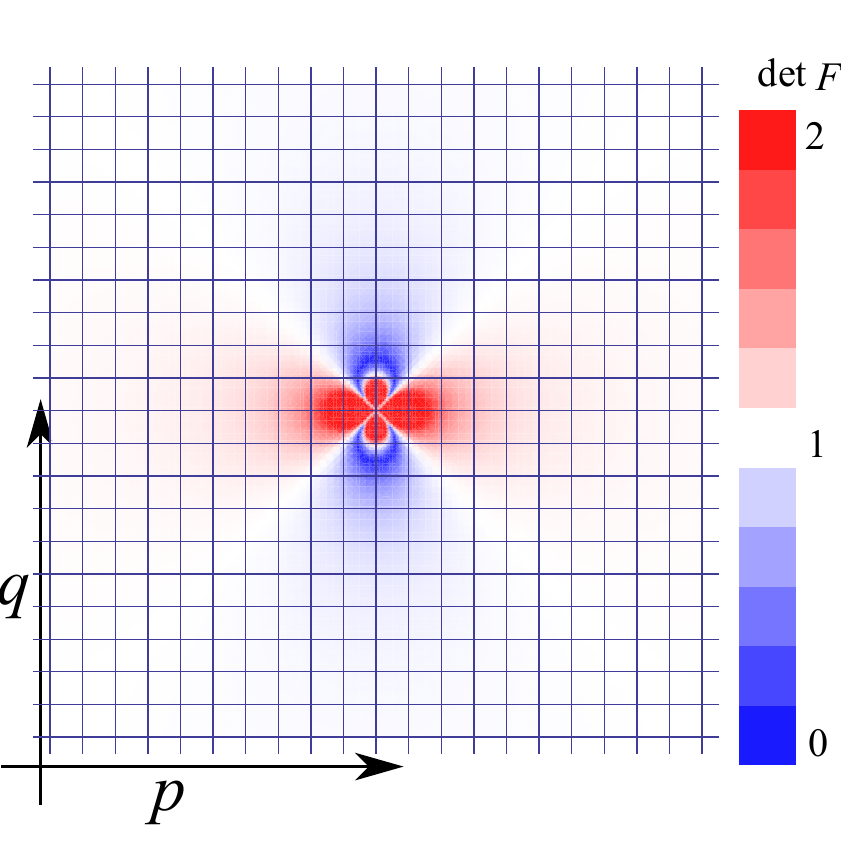}}
\subfigure[Fully non-linear solution]{\label{fullynonlinear}\includegraphics[width=0.24\textwidth]{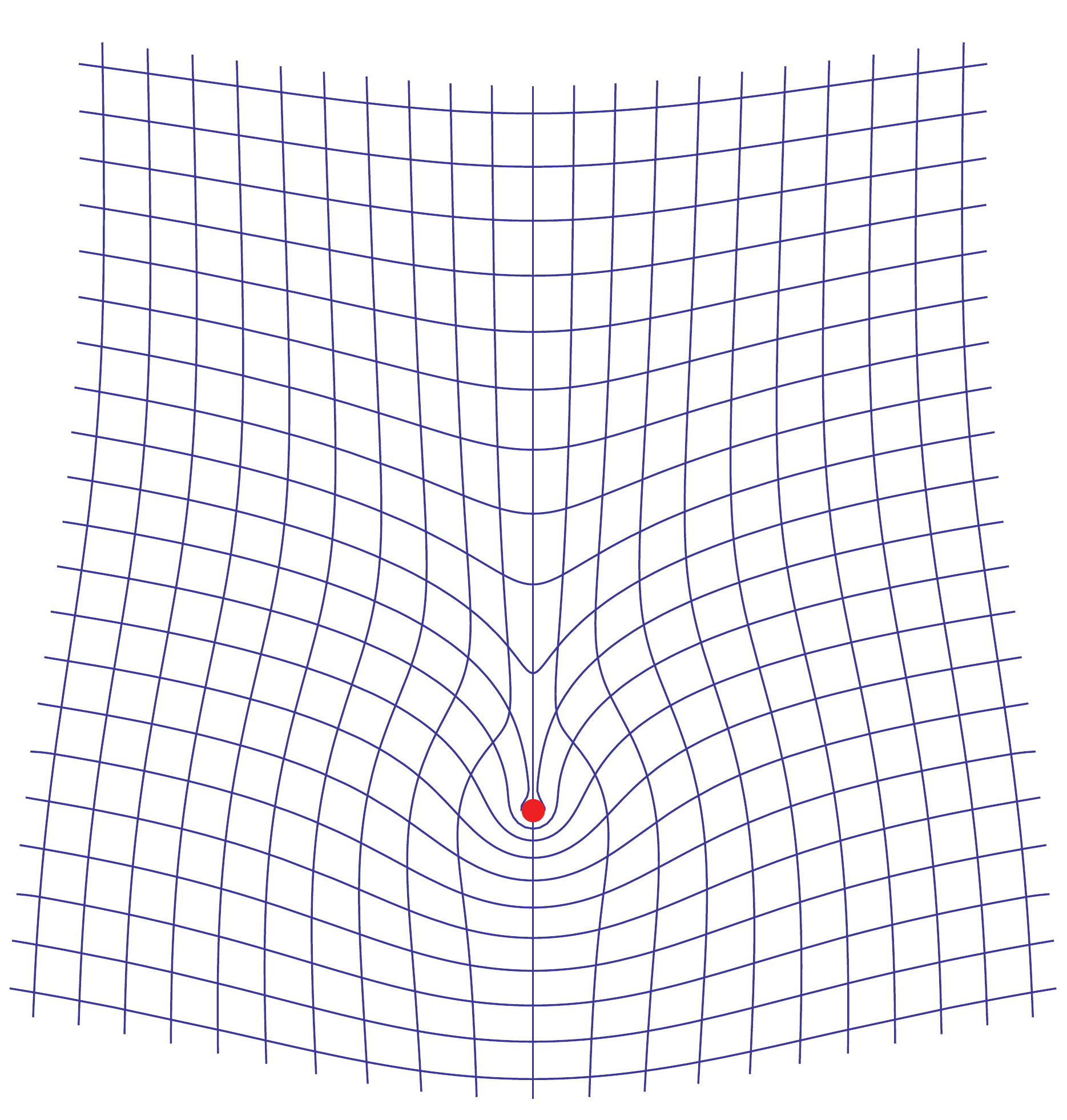}}
\caption{Deformation caused by a point force of magnitude $2 \mu l$ on a  2-D square grid with spacing $0.1 l$, where $l$ is an arbitrary length.  (a) Exactly area preserving solution given by eqn.\ (\ref{2dsol}). (b) Conventional ``incompressible'' Kelvin solution showing marked area changes near the point of application of the force. (c) Reference state colored by the area change caused by the Kelvin solution. (d) Numerical fully non-linear deformation of a material following eqn. (\ref{nlenergy}) with  $B=100\mu$. The force is applied to a rigid  disk of radius $0.025 l$.}\label{2dkelvin}
\end{figure*}

To resolve these shortcomings we first implement volume-conservation exactly by using a coordinate transformation originally deployed by Gauss but only recently introduced into elasticity\cite{carroll2007generating,ben2010swelling}. Linearizing the result leads to a linear theory of elasticity with perfect volume conservation that yields qualitatively different results than the theory that preserves volumes only to linear order. Consider a 2-D elastic reference state labeled with the  Cartesian coordinates $(p,q)$ and a target state labeled with the coordinates $(x,z)=(p,q)+\mathbf{u}$, as shown in fig.\ \ref{targandref}a. Conventionally the deformation between these states is described by  $x(p,q)$ and $z(p,q)$, so the   deformation gradient  is
\begin{equation}
F=\left(\begin{array}{cc} 
\frac{\partial x}{\partial  p}\big|_{q}&\frac{\partial  x}{\partial  q}\big|_{p}\\\frac{\partial  z}{\partial  p}\big|_{q}&\frac{\partial  z}{\partial  q}\big|_{p}\end{array}\right).
\end{equation}
However, if we describe the deformation via  a mixed coordinate system $z(x,q)$ and $p(x,q)$, as sketched in fig. \ref{targandref}b, we can implement the volume constraint exactly\cite{ben2010swelling}. To  understand this transformation, consider a patch of material specified by $q_1<q<q_2$ and $x_1<x<x_2$ shown in fig. \ref{targandref}b. In both the target and the reference state this patch has two parallel straight sides. Incompressibility requires the area $A$ of the patch to be the same in both  states:
\begin{equation}
A=\int_{q_1}^{q_2}[p(x_2,q)-p(x_1,q)]\mathrm{d}q=\int_{x_1}^{x_2}[z(x,q_2)-z(x,q_1)]\mathrm{d}x.
\end{equation}
If our patch is small in both dimensions, so that $q_2=q_1+\delta q$ and $x_2=x_1+\delta x$, we can evaluate these integrals to first order as $\delta x \delta q \frac{\partial p}{\partial x}\big|_{q}$ and  $\delta x \delta q \frac{\partial z}{\partial q}\big|_{x}$.  Equality then yields $ \frac{\partial p}{\partial x}\big|_{q}=\frac{\partial z}{\partial q}\big|_{x}$, a condition that is satisfied if we introduce a new field $\psi(x,q)$ such that 
\begin{equation}
p(x,q)=\frac{\partial \psi}{\partial q}\bigg|_x \mathrm{\ \ \ \ and  \ \ \ }  z(x,q)=\frac{\partial \psi}{\partial x}\bigg|_q.\label{introducepsi}
\end{equation}
The deformation gradient $F(x,q)$ in terms of the new functions $p(x,q)$ and $z(x,q)$ is given by
\begin{equation}
F=  \left(
\begin{array}{cc}
\left(  \frac{\partial p}{\partial x}\big|_{q}\right)^{-1} &  -\frac{\partial p}{\partial q}\big|_{x} \left(\frac{\partial p}{\partial x }\big|_{q}\right)^{-1} \\
 \frac{ \partial z}{\partial x}\big|_{q} \left(\frac{\partial p}{\partial x }\big|_{q}\right)^{-1} &\frac{\partial z}{\partial q}\big|_{x}-\frac{\partial p}{\partial q}\big|_{x} \frac{\partial z}{\partial x}\big|_{q}\left(\frac{\partial p}{\partial x}\big|_{q}\right)^{-1} \end{array}\label{streamF}
\right).
\end{equation}
Henceforth we regard all functions as functions of $x$ and $q$. Rewriting $F(x,q)$ in terms of $\psi$ yields 
\begin{equation}
F=\left(
\begin{array}{cc}
 \frac{1}{\psi _{{xq}}} & -\frac{\psi _{{qq}}}{\psi _{{xq}}} \\
 \frac{\psi _{{xx}}}{\psi _{{xq}}} & \psi _{{xq}}-\frac{\psi _{{qq}} \psi _{{xx}}}{\psi _{{xq}}}
\end{array}
\right),\label{2dF}
\end{equation}
which automatically satisfies $\Det{F}=1$.  We now  substitute this result into eqn.\ (\ref{tradnleqns}), at which point it is natural to regard $P$ and $\mathbf{f}$ also as functions of $x$ and $q$.    Starting from the undeformed state  $\psi=xq$, in the small strain limit, we write
\begin{align}
\psi(x,q)&=x q + \alpha(x,q)
\end{align} 
 where $\alpha \ll 1$, and further we assume that $|\nabla P| \ll 1$ . The first order expansion of $F$ and $F^{-T}$ are then
 \begin{equation}
F=\left(\begin{array}{cc}
 1-\alpha_{xq}& -\alpha_{qq} \\
\alpha_{xx} & 1+\alpha_{xq}
\end{array}
\right),
\end{equation}
\begin{equation}
F^{-T}=\left(\begin{array}{cc}
 1+\alpha_{xq}& -\alpha_{xx}  \\
\alpha_{qq}& 1-\alpha_{xq}
\end{array}
\right).\label{Flin}
\end{equation}
We now expand eqn.\ (\ref{tradnleqns})  to leading order in both $\alpha$ and $\nabla P$ to establish the governing equations of incompressible linear elasticity. To first order the partial derivative identities  $\frac{\partial}{\partial x}\big|_z=\frac{\partial}{\partial x}\big|_q$ and $\frac{\partial}{\partial z}\big|_x=\frac{\partial}{\partial q}\big|_x$ hold, and hence the expansion gives
\begin{equation}
\mu\left(
 \begin{array}{c}
-\alpha _{{qqq}}-\alpha _{{xxq}} \\
\alpha _{{xqq}}+\alpha _{{xxx}}
\end{array}
\right)-\left(
\begin{array}{c}
 P_x \\
P_q
\end{array}
\right)=-\mathbf{f}.\label{2deqilib}
\end{equation}
In a region with $\mathbf{f}=0$,  $\alpha$ is again biharmonic\cite{ben2010swelling}: $\alpha _{{qqqq}}+2 \alpha _{{xxqq}}+\alpha _{{xxxx}}=0 \label{biharmonicalpha}$. These equations are identical to those governing the traditional linear elastic equations, eq.\ (\ref{tradelasteqn}), with the important identification $p\to x$. Thus the truly incompressible response of an infinite 2-D medium to a point force of $\mathbf{f}=f \mathbf{\hat{q}}$ applied at the origin ($x=q=0$) is, by comparison with eqn.\ (\ref{2dsol}), 
\begin{align}
\alpha=-\frac{f x \log\left(q^2+x^2\right)}{8 \pi  \mu }\hspace{1em} P&=P_0+\frac{f q}{2 \pi   \mu(q^2+x^2) } \label{2dsolareapreserving}.
\end{align}
Similarly, any problem solved within the traditional streamline formalism can be imported into the strictly incompressible formalism with the  identification $p\to x$. However, one should not conclude the two  are equivalent, as we will now detail. Comparing the strictly incompressible and traditional solutions  in fig.\ \ref{2dkelvin}(a-c), we see that both  produce unphysical infinite displacement and self intersection near the point  force and are nearly identical far from the point of application of the force. However, there is a significant intermediate region where the Kelvin solution produces visible area changes, while the new solution does not. 

\begin{figure}
\includegraphics[width=0.4\textwidth]{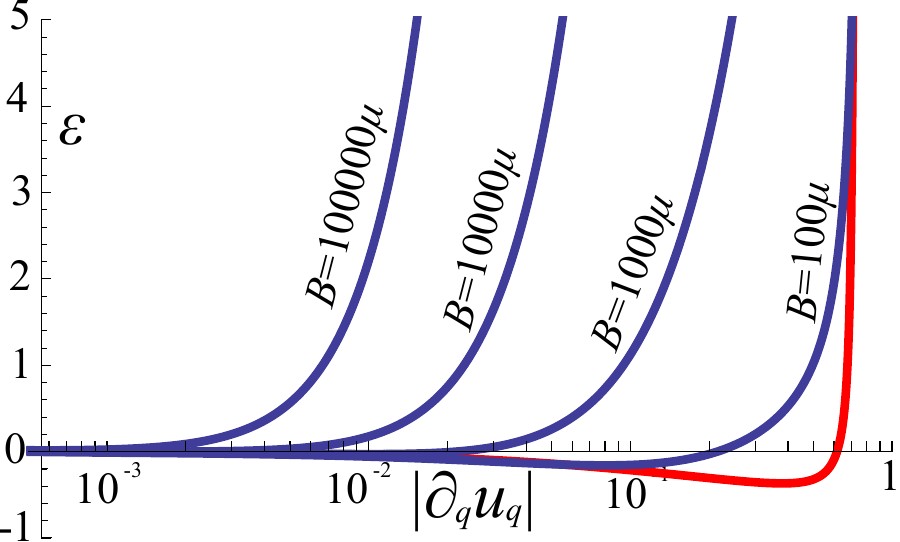}
\caption{Fractional energy discrepancy $\epsilon$ between the full non-linear finite element energy densities, $E_{nl}(F_{nl})$ (see eqn. (\ref{nlenergy})), and the approximate energy densities derived from the linear solutions, $E_{nl}(F)$, as a function of strain magnitude $|\partial_q u_q|$. The blue lines use the Kelvin ``incompressible'' linear solution,  fig. \ref{linear}, while the red line uses the exactly incompressible  solution, fig.\ \ref{streamline}. The data is taken along the material line $p=0$, $q<0$. The Kelvin solution only agrees with the full solution when strains are small compared to $\sqrt{\mu/B}$. The curve for the exactly incompressible solution is plotted using $B=100000\mu$, but  all modulus ratios considered here give essentially the same curve, showing the solution is works well for all materials with high bulk modulus untill strains approach one.}\label{2dkelvincompare}
\end{figure}

To assess the relative merits of the two solutions, we use the commercial finite element package ABAQUS to compute the fully non-linear response of a 2-d neo Hookean material  to a force $f$ concentrated on a small rigid line with a range of bulk moduli $B \gg \mu$, obeying  eqn.\ (\ref{nlenergy}). The elastic full-space is scale free but in terms of an arbitrary unit-length $l$, we assume that the force has magnitude $2\mu l$ and the line has radius $a=0.025 l$. The dimensionless number  $f/(\mu a)=80$ is thus large, meaning the force is point-like, being highly concentrated on a small disk. The rigid disk is embedded in the center of a material matrix of size $10000 l\times 10000l$. Due to the left-right symmetry, only the right half of the domain is simulated with symmetric boundary conditions on the left boundary and the other three boundaries being clamped, while a no-slip boundary condition is applied between the rigid line and the solid. We use a mesh of linear quadrilateral elements with  a total of $1641330$ degrees of freedom that has 100 evenly spaced elements along the half-length of the rigid line and that coarsens geometrically with distance from it.   In Fig.\ \ref{2dkelvin}(d) we depict one example of such a deformation, with $B/\mu=100$, and see that it is almost exactly area-preserving and lacks  the infinite displacement and self-intersection of the two linear solutions. We now compare the deformation gradient due to the linear solution, $F$ associated with the linearized incompressibility condition and the new exactly isochoric case, with the full non-linear solution, $F_{nl}$, at equivalent material points,  by computing the fractional discrepancy between the energy $\epsilon =[E_{nl}(F_{nl})-E_{nl}(F)]/E_{nl}(F_{nl})$.   Fig.\ \ref{2dkelvincompare} plots this fractional discrepancy $\epsilon$ as a function of the strain $\partial_q u_q$ for material points beneath the point of application of the force and for materials with a range of ratios $B/\mu$. We see that the traditional Kelvin solution only agrees with the full non-linear solution when the strain is small, i.e. $|\nabla u| \ll \sqrt{\mu/B}$, so as the materials become more incompressible, the traditional solution remains accurate only for smaller and smaller strains. In contrast, the exactly area preserving solution only breaks down at when $\partial_q u_q\sim1$, at which point geometric non-linearities dominate so that no linear theory can work.

A similar  approach can be implemented  in 3-D axisymmetric situations.  A deformation from a reference state described with the  polar-coordinates  $(\rho,\theta, q)$ to a target state described by  $(r, \phi,z)$, would usually be described by the functions $r(\rho,q)$ and $z(\rho,q)$, with axisymmetry requiring that $\phi=\theta$. We can implement traditional ``incompressibility'' ($\nabla \cdot \mathbf{u}=0$) by introducing a Stokes stream-function $\beta(r,q)$ such that $r=\rho-(1/\rho)\beta_q$ and $z=q+(1/r) \beta_\rho$. If instead we describe the deformation via   $\rho(r,q)$  and $z(r,q)$, we can enforce  $\Det{F}=1$ by introducing the scalar field $\chi(r,q)$ such that
\begin{equation}
z(r,q)=\frac{\chi_r}{ r} \mathrm{\ \ \ \ \ and \ \ \ \ \ \ }\rho(r,q)= \sqrt{2 \chi_q}.
\end{equation}
In this case, if there is no deformation $\chi=\half r^2 q$, so, to linearize, we write $\chi=\half r^2 q+\beta(r,q)$, where $\beta$ is small. Expanding to first order in $\beta$ we see that, with the identification $r\to\rho$, the displacements and deformations are algebraically identical to those with the conventional Stokes stream-line function.

\begin{figure*}\centering
\subfigure[Exact volume preservation]{\label{streamlineflatplates}
\includegraphics[width=0.34\textwidth]{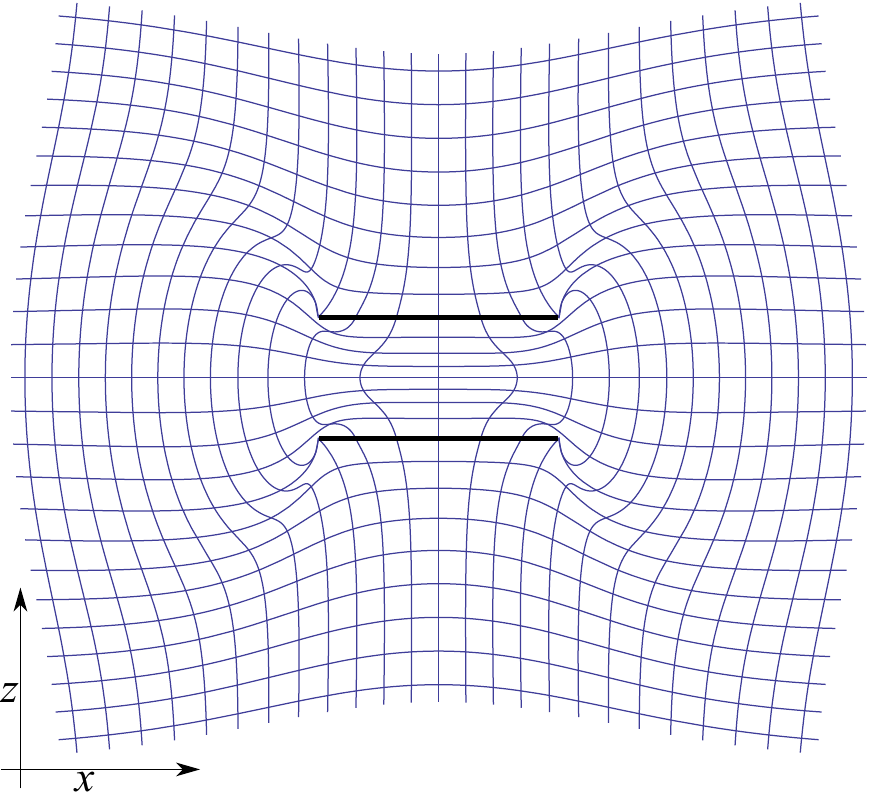}}
\subfigure[Conventional linear elasticity]{\label{linearflatplates}\includegraphics[width=0.4\textwidth]{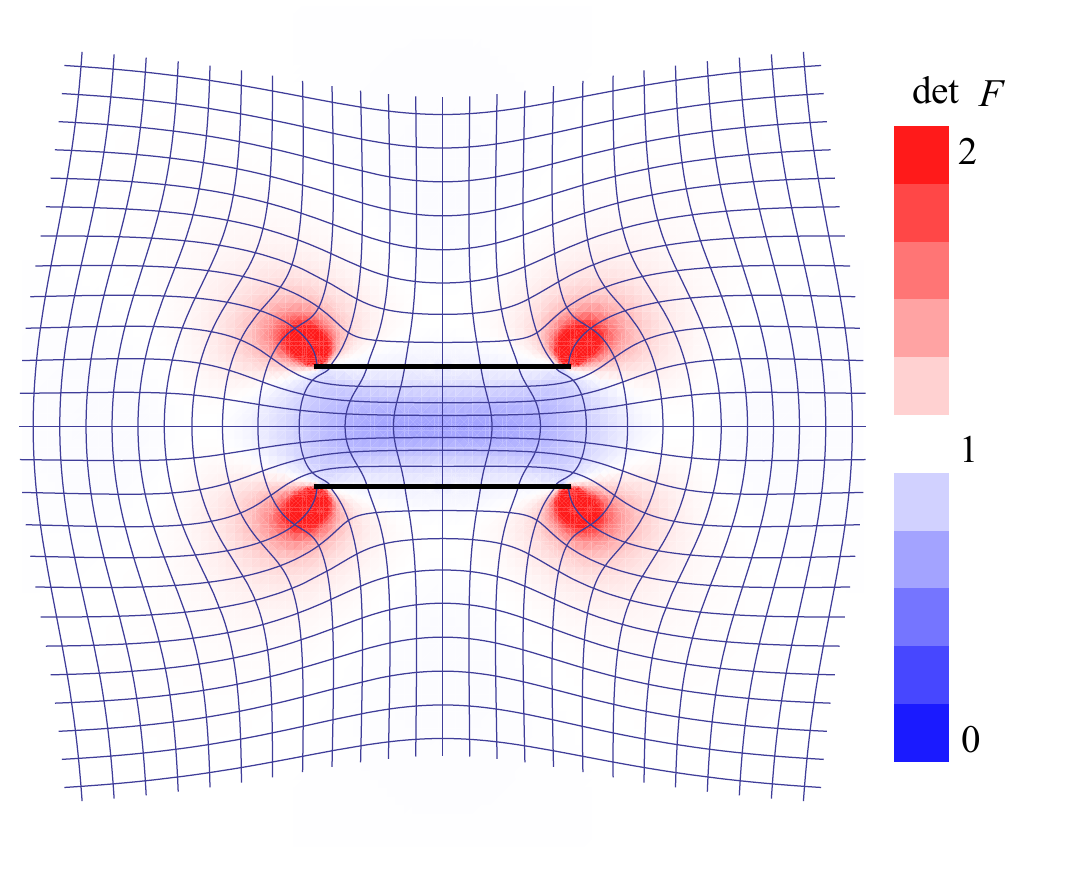}}
\caption{ Deformation caused by two parallel plates, with unit length and separation, embedded in a two dimensional elastic medium being pulled together. (a) is calculated using the exactly area preserving techniques outlined in this paper while (b) is calculated using traditional linear elasticity. The colour shows the (erroneous) fractional volume change in the traditional solution.}
\label{distload}
\end{figure*}

Our point force responses contain regions of divergent strain that are beyond any linear theory, so we close with an example where a distributed load leads to a finite response. Inspired by active-soft-matter systems such as magnetic-elastomers and muscular-tissues, we consider a two-dimensional elastic full-space containing two horizontal rigid plates, initially at  $q=\pm a$ and with width $2w$, that pull towards each other with a total force $f$.  The contribution to $\alpha$ from each plate is then given by integrating the point force solution along the plate weighted by a local force density $f(x_0)$, to get $\alpha=\int_{-w}^{w}f(x_0) x_0 \log\left(a^2+x_0^2\right)/({8 \pi  \mu }){\rm{d}} x_0$, where we must choose $f(x_0)$ so that it integrates to $f$ and has the correct profile to keep the plates flat. For a single plate, this would entail $f^{(q)}(x_0)=1/\sqrt{1-x_0^2}$ \cite{johnson1987contact}. For two plates we were unable to find a closed from solution for $f^{(q)} (x_0)$ but achieved almost perfect rigidity by taking the form  $f^{(q)}(x_0)=(A+ B x_0^2+ C x_0^4)/\sqrt{1-x_0^2}$ and fitting $A$, $B$ and $C$. We compare this fully area preserving solution to the traditional-elastic one in fig.\ \ref{streamlineflatplates}-\ref{linearflatplates}, and observe that  the traditional solution suffers substantial erroneous area loss between the plates. Although neither  solution solves  the full non-linear problem, our deformation field offers an admissible solution for consideration in a non-linear theory --- its fully non-linear energy can be evaluated and it can serve as a starting point in an attempt, either numerically or analytically, to minimize the fully non-linear energy. A detailed account that builds on our 2d analysis of exactly isochoric deformations of soft solids  documenting all the major exactly incompressible bulk and surface Green's functions for 2-D and axisymmetric elastic full and half spaces with and without large pre-strains is available here \cite{bigginsappendix}.

A natural objection at this point might be  that  it is inconsistent to linearize our theory and satisfy incompressibility perfectly, and that the new perspective thus adds nothing to the traditional stream-line formalism. However, we see considerable value in perfect volume preservation. Firstly, as shown above, it extends the range of validity of the linear theory from strictly infinitesimal strains to strains approaching unity. Secondly, it produces fields that can be substituted into the full non-linear energy as a starting point for building energy-estimates, numerical minimization or rigorous bounds. It thus provides a bridge connecting linear  and non-linear elasticity in incompressible systems. Thirdly, it will be a useful approximate method in contexts where the full non-linear problem is intractable and incompressibility is critical to understanding the key physics.  Finally, our approach may be useful in developing numerical elasticity and elastodynamic integrators for incompressible systems. Our approach is  analogous to the use of symplectic integrators that have become commonplace in classical dynamics. These also introduce a different representation of a problem that guarantees the exact preservation of key conserved quantities (often energy or momentum), and their use in numerics leads to stabler, more robust and faster simulations. In the elasticity of compressible materials, numerically one is often stymied by having to take an asymptotically small time-step because the speed of sound diverges. A scheme based on our method would allow the implementation of strict incompressibility with a finite time-step. It is by thus linking the linear and non-linear theories of incompressible soft solids that we anticipate the theory outlined in here will find application.


\end{document}